# Leveraging Large Language Models for Preliminary Security Risk Analysis: A Mission-Critical Case Study

Matteo Esposito
m.esposito@ing.uniroma2.it
University of Rome "Tor Vergata"
Rome, Lazio, ITA

Francesco Palagiano
palagiano.francesco@multitelsrl.it
Multitel di Lerede Alessandro & C. s.a.s.
Rome, Lazio, ITA

## ABSTRACT

Preliminary security risk analysis (**PSRA**) provides a quick approach to identify, evaluate and propose remediation to potential risks in specific scenarios. The extensive expertise required for an effective PSRA and the substantial ammount of textual-related tasks hinder quick assessments in mission-critical contexts, where timely and prompt actions are essential. The speed and accuracy of human experts in PSRA significantly impact response time. A large language model can quickly summarise information in less time than a human. To our knowledge, no prior study has explored the capabilities of fine-tuned models (**FTM**) in PSRA. Our case study investigates the proficiency of FTM to assist practitioners in PSRA. We manually curated 141 representative samples from over 50 mission-critical analyses archived by the industrial context team in the last five years. We compared the proficiency of the FTM versus seven human experts. Within the industrial context, our approach has proven successful in reducing errors in PSRA, hastening security risk detection, and minimizing false positives and negatives. This translates to cost savings for the company by averting unnecessary expenses associated with implementing unwarranted countermeasures. Therefore, experts can focus on more comprehensive risk analysis, leveraging LLMs for an effective preliminary assessment within a condensed timeframe.

## CCS CONCEPTS

• **Security and privacy** → *Security requirements*; *Domain-specific security and privacy architectures*; • **Applied computing** → **IT governance**; **Cyberwarfare**; • **Computing methodologies** → **Natural language generation**; **Information extraction**.

## KEYWORDS

Preliminary, Security, Risk, Management, Analysis, Large Language Model, LLM, Generative AI, ISO, Standards, Human Experts, Fine-Tuning

## 1 INTRODUCTION

In the realm of Information Technology (**IT**) security, organizations face the challenge of safeguarding digital assets against evolving threats [15, 5, 11]. To address the challenges, national governments and international entities, including the ISO, offer guidelines and regulations tailored to quality and risk analysis in mission-critical contexts (MCC), emphasizing the importance of ensuring the success and reliability of operations [5].

Risk analysis (**RA**) focus on comprehending the nature of risk and its characteristics, including, where appropriate, the risk level. RA involves a detailed consideration of uncertainties, risk sources, consequences, likelihood, events, scenarios, controls and their effectiveness [8]. Preliminary Risk Analysis (**PRA**), a.k.a. Preliminary Hazard Analysis, is used to quickly identify potential security risks relating to an IT system and its interfaces. PRA assesses the probability of occurrence and the severity of potential risks, aiming to propose solutions that effectively reduce, control, or eliminate them [12].

In our context, preliminary security risk analysis (**PSRA**) applies PRA principles to cyber-physical MCC, providing a structured approach to identifying, evaluating and remediating potential risks to achieve reliability of operations [15, 11]. PSRA is crucial to develop countermeasures to detect risks in aerospace, health care, defence, finance and emergency services, where any deviation from optimal performance significantly impacts the context [6].

More specifically, our PSRA in MCC centres on evaluating whether a specific scenario contains or does not contain a security risk without identifying the specific risk type. The text heavy nature of PSRA and the years of expertise required for a quick and effective analysis require a tool capable of rapidly learning and analyzing data, possibly outperforming human proficiency.

Large Language Models (**LLMs**) are designed to process and produce text that closely resembles human language. LLMs are trained on vast amounts of text to perform language tasks like translation, summarisation, question-answering, and text completion [2]. Therefore, LLMs are potentially valuable for PSRA. Our study contributions are two-fold: we present the first case study on leveraging LLMs in PSRA and evaluate the proficiency of the fine-tuned model (**FTM**), comparing it with seven human experts in risk analysis with different professional backgrounds.

Our findings are based on the experience and data acquired during the late summer of 2023, encompassing the initial implementation of our industrial context concept. Notably, the GPLLM exhibited suboptimal performance when compared to human experts. Conversely, the FTM surpassed six of seven human experts in proficiency and speed.

**Paper Structure:** Section 2 describes the study design. Section 3 presents the obtained results, and Section 4 discusses them. Section 5 highlights the threats to the validity of our study, and Section 6 draws the conclusion.

## 2 METHODOLOGY

Our empirical study was designed as a case study following established guidelines [10]. In the upcoming sections, we detail the specific research questions and goals that drive our study and the procedures we used for data collection and analysis.

### 2.1 Goal and Research Questions

We use the Goal Question Metric (GQM) approach [1] to formalise our goal as follows:

*Investigate* a fine-tuned model, *for the purpose of* evaluation, *with respect to* its proficiency in PSRA, *from the point of view of* practitioners, *in the context of* mission critical IT security.

Based on our goal, we defined two Research Questions ($RQ_s$).

> $RQ_1$ Can an LLM perform PSRA?

General purpose large language models (**GPLLMs**) assist humans in everyday tasks, from creative writing to data analysis [17, 2]. As GPLLMs, we selected OpenAI's *gpt-3.5-turbo-1106* model. Despite the broad capabilities of GP models, fine-tuning becomes essential for tasks requiring specific domain expertise [7]. For instance, Yang, Tang, et al. [16] achieved success by fine-tuning LLAMA on a manually curated financial dataset, enhancing the effectiveness of financial professionals. Moreover, Tufano et al. [14] utilized fine-tuning in the development of AthenaTest, an automated approach for generating unit test cases. Nevertheless, to the best of our knowledge, no previous study has explored the potential of FTM for preliminary security risk analysis. Therefore, investigating the proficiency of an FTM in this context enables practitioners to promptly identify security threats within the tested environment, particularly in mission-critical scenarios. After establishing the effectiveness of the model, we ask:

> $RQ_2$ Can an LLM outperform human expert?

The level of human expertise and years spent in the field directly impact PSRA quality and proficiency. Table 1 provides an overview of the experts' profiles. We can measure the time required for a new team member to attain the proficiency level of a senior team member in years or even decades. In contrast, LLMs can rapidly assimilate years of training within mere minutes. Therefore, it is essential to investigate the FTM proficiency in terms of accuracy and time, comparing it to human experts. We provided 40 samples to the FTM and seven human experts. We measured the proficiency of both human experts and the FTM by measuring their timing and computing metrics such as Accuracy, Precision, Recall, and F1-score [2].

### 2.2 Industrial context of the study

The context of our case study is an Italian company that has been operating in the civil and military security sector for over 30 years. It is dedicated to researching and developing new technologies for information security and provides products and services aimed at safeguarding data, both at rest and in motion. Additionally, it conducts design, verification, implementation, and certification interventions for security, including "What-if" Analysis.

### 2.3 Sample Selection

The sample selection derives from previously finalised risk analyses, thus enabling us to obtain the ground truth necessary for fine-tuning and evaluating the model. We engaged with the company Risk Analysis and Management Team (**RAMT**) and randomly selected excerpts from the existing finalised documents. To ensure the replicability of our findings, we excluded all sensitive information and samples that could compromise data anonymity. The final dataset comprises 141 samples, encompassing over 50 mission-critical analyses conducted over the last five years. The authors and the RAMT reviewed each sample and agreed unanimously on each classification.

### 2.4 Study setup and data collection

This section delineates our data collection methodology. Our study entailed analysing excerpts from PSRA interviews. The RAMT provided 200 samples from previously finalised PSRAs. We excluded 59 samples due to sensitive data, representativity concerns, and the need for data anonymity. Regarding representativity, our goal was to fine-tune the model with diverse scenarios, avoiding anchoring towards specific keywords or scenario descriptions. The final dataset comprised 141 samples, with 100 samples allocated for training and validation and the remaining 41 samples designated for testing.

OpenAI API's documentation states that fine-tuning requires providing conversation samples in a JSONL-formatted document[1]. When fine-tuning a model, three roles are available: the *system* role guides the model on how to behave or respond. In our case, we directed the model to reply in a JSON-like output to facilitate automated analysis. We asked the model to respond briefly with only three possible answers in the result field: **yes** if it identified a potential security threat, **no** otherwise, and **more** in case the user-provided insufficient information for a proper PSRA. Moreover, we allowed the model to provide a more open reply in the message field of the custom-defined JSON for future works. The *user* role simulates the end user asking a question. Similarly to how we label instances in machine learning, the *assistant* role simulates the model to provide examples of correct answers. We used 70 samples as a training set and 30 as a validation set.

Finally, we tasked the model and the seven human experts with analyzing the last 41 samples (**testing samples**). We instructed the experts to classify the samples following the model's three possible answers, i.e., yes, no, and more.

---

[1] https://platform.openai.com/docs/guides/fine-tuning

Table 1: Expert Profiles

| Expert | Experience (Years) | Expertise | Common Role | Professional Level | Corporate Level | Age Range |
|---|---|---|---|---|---|---|
| 1 | 35 | IT & Telecom | Security Secretariat | Employee | Manager | 51-60 |
| 2 | 5 | Logistics | Cipher Operator | Employee | Senior | 31-40 |
| 3 | 45 | IT & Telecom | Director of Security | Executive | Director | 81-90 |
| 4 | 26 | Logistics | Coordinator of Interventions | Manager | Manager | 51-60 |
| 5 | 5 | Logistics | Cipher Operator | Employee | Senior | 41-50 |
| 6 | 3 | IT & Telecom | System Administrator | Employee | Junior | 21-30 |
| 7 | 15 | IT & Telecom | Security Functionary | CEO | Owner | 41-50 |

## 2.5 Data Analysis

This section presents the data analysis procedure we employed in addressing our research questions. To answer $RQ_1$, we fine-tuned the selected GPLLM, acting as the baseline, obtained our FTM, and evaluated both on the testing samples. For $RQ_2$, we conducted an evaluation involving seven human experts on the same testing samples and compared their performance against the FTM. The assessment covered multiclass classification carried out by the models and the human experts, employing widely used metrics: Accuracy, Precision, Recall, and F1-Score. In RA not all misclassifications are deemed equal [13], and their significance can vary from one context to another. In our context, a "yes" instead of a "no" is less severe than a "no" instead of a "yes" or a "no" instead of a "more". To account for the varying severity of errors, we weighted the metrics. We utilized ScikitLearn (**SL**)[2] to calculate accuracy metrics. Specifically, we used "weighted" as the *average* parameter for weighted evaluation and "micro" for unweighted evaluation. When specifying "micro", SL computes metrics globally, counting total true positives, false negatives, and false positives across all classes. It then calculates accuracy metrics using these global counts, thus giving equal weight to each instance. It's important to note that when we mention the "*proficiency*" of a model or a human expert, we consider accuracy metrics, the number of errors, and the time taken to evaluate the samples.

## 2.6 Replicability

Our replication package includes a Python notebook importable into Google Colab with fine-tuning data, questionnaire answers, error type weights, and raw model responses. The package is available on Zeondo: https://zenodo.org/doi/10.5281/zenodo.10501335.

## 3 RESULTS

### 3.1 $RQ_1$ Can an LLM perform PSRA?

Table 2 presents the proficiency comparison between human experts and LLMs with the 'micro' and 'weighted' average types. According to Table 2, FTM consistently outperforms the baseline, i.e., GPLLM, in each accuracy metric. Moreover, FTM exhibits high precision in both average types, suggesting low rates of false positives. Furthermore, FTM achieves a weighted recall of 0.8814, suggesting it can effectively discover preliminary security risks with a low rate of false negatives. **This result suggests that FTM can effectively perform PSRA**.

[2]https://scikit-learn.org/stable/modules/generated/sklearn.metrics.precision_recall_fscore_support.html

### 3.2 $RQ_2$ Can an LLM outperform human expert?

According to Table 2, human experts outperform the GPLLM in each accuracy metric and the overall number of incorrectly classified samples (i.e., Errors #). Nonetheless, FTM outperforms six of seven human experts in all accuracy metrics, in the number of errors, and in analysis time. **This result suggests that FTM can outperform human experts in PSRA.**

Furthermore, despite Expert 4 (E4) reporting the highest number of incorrectly classified samples (i.e., 14), the weighted precision and recall scores are higher than those of, for instance, E5, highlighting that errors from E5 were more severe than those from E4. **These results suggest that weighting error types in PSRA is essential for result interpretation**.

## 4 DISCUSSIONS

Regarding $RQ_1$, Table 2 shows that although the GPLLM performed poorly, it was still able to correctly identify 47% of the samples in less than a minute. This outcome aligns with the inherent advantages LLMs have over human speed in analyzing data. Therefore, we highlight that **fine-tuning on a small dataset can significantly enhance proficiency** [9, 18], resulting in improvements for the weighted metrics of up to 204% in Recall and 314% in F1.

Regarding $RQ_2$, more experienced experts lead to less severe errors. E3 exemplifies that years of experience make a difference in the human world. On the other hand, in 18 seconds, the FTM challenged the accuracy of 45 years of experience. Nevertheless, our case study aimed not to replace human experts but to investigate a preliminary tool for quickly scanning a context and detecting preliminary security risks. **Experts should leverage the FTM to enhance their analysis capabilities, and to focus to the more extensive RA**.

## 5 THREATS TO VALIDITY

In this section, we discuss the threats to the validity of our case study. We categorised the threats in Conclusion, Internal, Construct and External validity following established guidelines [4].

**Conclusion Validity** focuses on how we draw conclusions based on the design of the case study, methodology, and observed results [4]. Our conclusions rely on the specific accuracy metrics chosen, and there may be other aspects or dimensions of performance that we did not consider. To address this potential limitation, we selected metrics from recent related studies that have faced the challenge of validating FTM proficiencies in specific tasks [2].

Table 2: Proficiency comparison between human experts and LLMs.

| Metric | Average Type | E1 | E2 | E3 | E4 | E5 | E6 | E7 | FTM | GPLLM |
|---|---|---|---|---|---|---|---|---|---|---|
| Accuracy | micro | 0.8 | 0.8 | 1 | 0.65 | 0.6 | 0.8 | 0.75 | 0.9 | 0.525 |
|  | wheighted | 0.8 | 0.8 | 1 | 0.65 | 0.6 | 0.8 | 0.75 | 0.9 | 0.525 |
| Precision | micro | 0.8 | 0.8 | 1 | 0.65 | 0.6 | 0.8 | 0.75 | 0.9 | 0.525 |
|  | wheighted | 0.751 | 0.751 | 1 | 0.473 | 0.225 | 0.752 | 0.680 | 0.881 | 0.463 |
| Recall | micro | 0.8 | 0.8 | 1 | 0.65 | 0.6 | 0.8 | 0.75 | 0.9 | 0.525 |
|  | wheighted | 0.593 | 0.615 | 1 | 0.433 | 0.369 | 0.593 | 0.517 | 0.818 | 0.269 |
| F1 Score | micro | 0.8 | 0.8 | 1 | 0.65 | 0.6 | 0.8 | 0.75 | 0.9 | 0.525 |
|  | wheighted | 0.574 | 0.614 | 1 | 0.380 | 0.278 | 0.561 | 0.456 | 0.813 | 0.196 |
| Time (# Errors) |  | 8m 12s (8) | 9m 24s (8) | 12m 51s (0) | 15m 25s (14) | 14m 4s (9) | 11m 35s (8) | 6m 52s (10) | 18s (4) | 1m (19) |

**Internal Validity** is the extent to which an experimental design accurately identifies a cause-and-effect relationship between variables [4]. Our study relies on 141 samples, which can potentially be biased from the sample selection and the MCC. We addressed this issue by sampling over 50 mission-critical analyses conducted by the industrial context team on different fields, from national security to health and education.

**Construct validity** concerns how our measurements reflect what we claim to measure [4]. Our specific design choices, including our measurement process and data filtering, may impact our results. To address this threat, we based our choice on past studies and used well-established guidelines in designing our methodology [10, 1].

**External validity** concerns how the research elements (subjects, artefacts) represent actual elements [4]. Our case study focused on an Italian company operating in the civil and military security field. The use of the Italian language and the specific characteristics of this company may limit the generalizability of the findings to other organisations or fields. We addressed this concern similar to the internal validity threats by sampling from over 50 mission-critical risk analyses across various fields. Moreover, we chose a GPLLM with no specific Italian language advantages or restrictions. Given the inherent language capabilities of the model and the context-less nature of RA, having the samples in Italian should not pose any generalizability issues [9, 3].

## 6 CONCLUSIONS

In this section, we briefly draw our conclusions. Our study delved into the proficiency of LLMs in PSRA. The findings emphasized that GPLLMs are suitable for PSRA, but fine-tuning is essential to narrow down the model's focus towards specific tasks. Notably, our FTM outperformed six out of seven human experts in both accuracy metrics, the number of errors, and evaluation time. Moreover, FTM misclassifications were less severe than those made by their human counterparts, as highlighted by the weighted metrics. Hence, suggesting that FTM can be used as a 'copilot' in the decision making process of PSRA.

In the industrial context, our approach has effectively reduced errors in PSRA, accelerated risk detection, and minimized both false positives and, most importantly, false negatives, which are critical for the reliability of operations. Consequently, our approach prevented unnecessary costs for the company, avoiding expenses incurred in implementing unnecessary countermeasures (i.e., false positives). As a result, the RAMT can now concentrate on the more extensive risk analysis, obtaining an effective preliminary analysis in a short period by leveraging our FTM. Therefore, the key benefit of the FTM is to support the decision-makers in focusing, quickly, on the vulnerabilities that really need to be addressed without concerns about incorrect indications on the items to be worked on.

Our current research efforts are focusing on leveraging retrieval augmented generation to further enhance the model adding more risk analysis frameworks, to allow a cross-evaluation between different reference standards, specific threat detection, personnel cost, and remediation estimation for a more comprehensive security risk analysis.